\documentclass[12pt]{article}
\usepackage[dvips]{color}
\usepackage{epsfig}
\usepackage{amsmath}
\usepackage{graphicx}
\def\Box{\hbox{$\rlap{$\sqcup$}\sqcap$}}
\begin{document}
\title{\bf Bouncing Universe with Non-minimally Coupled Quintom Matter}
\author{{M. R. Setare $^{a}$ \thanks{Email: rezakord@ipm.ir}\hspace{1mm}
, J. Sadeghi $^{b}$\thanks{Email: pouriya@ipm.ir}\hspace{1mm} and A.
Banijamali$^{b}$ \thanks{Email: abanijamali@umz.ac.ir}}\\
{ $^{a}$Department of Science, Payame Noor University, Bijar, Iran
}\\ { $^b$Sciences Faculty, Department of Physics, Mazandaran
University, }\\{ P .O .Box 47415-416, Babolsar, Iran} }

\maketitle

\begin{abstract}
In this letter, we study the condition for a generalized DBI action
providing a quintom scenario of dark energy. We consider a
development of string-inspired quintom by introducing non-minimal
coupling. Then we show that the bouncing solution can appear in the
universe dominated by the non-minimally coupled quintom matter.

\end{abstract}
\newpage
\section{Introduction}
Bouncing universes are those that go from an era of accelerated
collapse to an expanding era without displaying a singularity. Due
to this fact that the initial singularity may be absent in realistic
descriptions of the universe, many cosmological solutions displaying
a bounce were examined in the last decades (for a very recent review
see \cite{rev}). According to the brane-world scenarios of
cosmology, if the bulk space is taken to be an AdS black hole with
charge, the universe can bounce \cite{39}. That is, the brane makes
a smooth transition from a contracting phase to an expanding phase.
From a four-dimensional point of view, singularity theorems
\cite{40} suggest that such a bounce cannot occur as long as certain
energy conditions apply. Hence, a key ingredient in producing the
bounce is the fact that the bulk geometry may contribute a negative
energy density to the effective stress-energy on the brane
\cite{41}. At first sight these bouncing brane-worlds are quite
remarkable, since they provide a context in which the evolution
evades any cosmological singularities while the dynamics is still
controlled by a simple (orthodox) effective action. In particular,
it seems that one can perform reliable calculations without
deliberation on the effects of quantum gravity or the details of the
ultimate underlying theory. Hence, several authors \cite{{42},{43}}
have pursued further developments for these bouncing brane-worlds.\\
Nowadays it is strongly believed that the universe is experiencing
an accelerated expansion. Recent observations from type Ia
supernovae \cite{SN} in associated with Large Scale Structure
\cite{LSS} and Cosmic Microwave Background anisotropies \cite{CMB}
have provided main evidence for this cosmic acceleration. In order
to explain why the cosmic acceleration happens, many theories have
been proposed. Although theories of trying to modify Einstein
equations constitute a big part of these attempts, the mainstream
explanation for this problem, however, is known as theories of dark
energy. It is the most accepted idea that a mysterious dominant
component, dark energy, with negative pressure, leads to this cosmic
acceleration, though its nature and cosmological origin still remain
enigmatic at present.\\
In modern cosmology of dark energy, the equation of state parameter
(EoS) $\omega=\frac{p}{\rho}$ plays an important role, where $p$ and
$\rho$ are its pressure and energy density, respectively. To
accelerate the expansion, the EoS of dark energy must satisfy
$\omega<-\frac{1}{3}$. The simplest candidate of the dark energy is
a tiny positive time-independent cosmological constant $\Lambda$,
for which $\omega=-1$ \cite{{Einstein:1917},{cc}}. However, it is
difficult to understand why the cosmological constant is about 120
orders of magnitude smaller than its natural expectation (the Planck
energy density). This is the so-called cosmological constant
problem. Another puzzle of the dark energy is the cosmological
coincidence problem: why are we living in an epoch in which the dark
energy density and the dust matter energy are comparable?. As a
possible solution to these problems various dynamical models of dark
energy have been proposed, such as quintessence \cite{{c6},{c7}}.
The analysis of the properties of dark energy from recent
observations mildly favor models with $\omega$ crossing -1 in the
near past. So far, a large class of scalar-field dark energy models
have been studied, including tachyon \cite{tachyon}, ghost
condensate \cite{ghost1,ghost2} and quintom
\cite{{c9},{c12},{c14},{c15}}, and so forth.  In addition, other
proposals on dark energy include interacting dark energy models
\cite{intde}, braneworld models \cite{brane}, and holographic dark
energy modeles \cite{holo}, etc. The Ref.\cite{c9}  is the first
paper showing explicitly the difficulty of realizing $\omega$
crossing over -1 in the quintessence and phantom like models.
Because it has been proved \cite{{c9},{c10},{c11}} that the dark
energy perturbation would be divergent as the equation of state
$\omega$ approaches to -1. The quintom scenario of dark energy is
designed to understand the nature of dark energy with $\omega$
across -1. The quintom models of dark energy differ from the
quintessence, phantom and k-essence and so on in the determination
of the cosmological evolution and the fate of the universe.
\\To realize a viable quintom scenario of dark energy it
needs to introduce extra degree of freedom to the conventional
theory with a single fluid or a single scalar field. The first model
of quintom scenario of dark energy is given by Ref.\cite{c9} with
two scalar fields. This model has been studied in detail later on
\cite{{c12},{c14},{c15}}. Recently there has been an upsurge in
activity for constructing such model in string theory \cite{c16}. In
the context of string theory, the tachyon field in the world volume
theory of the open string stretched between a D-brane and an
anti-D-brane or a non-BPS D-brane plays the role of scalar field in
the quintom model \cite{c17}. The effective action used in the study
of tachyon cosmology consists of the standard Einstein-Hilbert
action and an effective action for the tachyon field on unstable
D-brane or D-brane anti D-brane system. What distinguishes the
tachyon action from the standard Klein- Gordon form for scalar field
is that the tachyon action is non-standard and is of the "
Dirac-Born-Infeld " form \cite{c23, c18}. The tachyon potential is
derived from string theory itself and has to satisfy some definite
properties to describe
tachyon condensation and other requirements in string theory.\\
In this paper, we consider an action for tachyon non-minimally
coupled to gravity \cite{c19} inspired by the string theory.  Then
we study the bouncing solution in the universe dominated by
non-minimally coupled quintom matter.

\section{Bouncing behaviour of non-minimally coupled tachyon gravity with extra term}
We will start with a detailed examination on the necessary
conditions required for a successful bounce \cite{bou}. During the
contracting phase, the scale factor $a(t)$ is decreasing, i.e.,
$\dot{a} < 0$, and in the expanding phase we have $\dot{a} > 0$. At
the bouncing point,$\dot{a}=0$, and around this point $\ddot{a} > 0$
for a period of time. Equivalently in the bouncing cosmology the
hubble parameter H runs across zero from $H < 0$ to $H > 0$ and $H =
0$ at the bouncing point. A successful bounce requires around this
point,
\begin{equation} \label{boun}
\dot{H}=\frac{-1}{2M_{P}^{2}}(\rho+P)=\frac{-1}{2M_{P}^{2}}\rho
(1+\omega)>0 \end{equation}
  where $M_{P}=\frac{1}{\sqrt{8\pi G}}$.\\
  Now we
consider the action of Ref.\cite{c20} for tachyon non-minimally
coupled to gravity, then we add an extra term $T \Box T$ to the
usual terms in the square root of this action. In that case the
following action is the same as Ref.\cite{c21} just different to the
$ Rf(T),$
\begin{equation} \label{ac}
S=\int d^{4}x \sqrt{-g} \left[\frac{M_{P}^{2}}{2}Rf(T) -
AV(T)\sqrt{1-\alpha'g^{\mu\nu}\partial_{\mu}T\partial_{\nu}T+\beta'T
\Box T}\right],
\end{equation}
where $f(T)$ is a function of the tachyon $T$ and corresponds to the
non-minimal coupling factor. Here $V(T)$ is the tachyon potential
which is bounded and reaching its minimum asymptotically.
\\
Action (\ref{ac}) generalizes the usual "Born- Infeld- type" action
for the effective description of tachyon dynamics which can be
obtained by the stringy computations for a non- BPS D3- brane in
type II theory. The extra term in action (\ref{ac}) has a
significant cosmological consequence, so we cannot exclude its
existence in an action such as
the "Born- Infeld- type" action.\\
The model with operator $T \Box T$ for realizing of $\omega$
crossing -1 has been proposed in \cite{c12}. The operator  $T \Box
T$ can be rewritten as a total derivative term which makes no
contribution after integration and a term which renormalizes the
canonical kinetic term as has discussed in \cite{c21}. So, if one
consider a renormalizable Lagrangian, the operator  $T \Box T$ will
not be included. Ref.\cite{c12} considered a dimension-6 operator as
$ (\Box T)^{2}$. However in the present paper, the operator $T \Box
T$ appears at the same order as the operator
$\partial_{\mu}T\partial^{\mu}T$ does in the "Born- Infeld- type"
action and also we take into account scalar curvature non-minimally coupled to the tachyon field.\\
The action (\ref{ac}) can be brought to the simpler form to derive
the equation of motion, energy density and pressure, by performing a
conformal transformation as follows:
\begin{eqnarray}
g_{\mu\nu}\longrightarrow f(T) g_{\mu\nu}.
\end{eqnarray}
The above conformal transformation yields to the following action:\\
$$S=\int d^{4}x\sqrt{-g}\
\Bigg[\frac{M_{P}^{2}}{2}(R-\frac{3}{2}\frac{f'^{2}}{f^{2}}\partial_{\mu}T\partial^{\mu}T)$$

\begin{eqnarray}
-A\tilde{V}(T)\sqrt{1-(\alpha'f(T)-2\beta'f'(T)T)\partial_{\mu}T\partial^{\mu}T+\beta'f(T)T
\Box T}\Bigg]\,
\end{eqnarray}
where $\tilde{V}(T)=\frac{V(T)}{f^{2}}$ is now the effective
potential of the tachyon.\\
 For a flat Friedman- Robertson- Walker
(FRW) universe and a homogenous scalar field $T$, the equation of
motion can be solved equivalently by the following two equations,
$$\ddot{\psi}+3H\dot{\psi}=(\frac{2\beta'f'T-\alpha'f}{fT})\dot{\psi}\dot{T}-\frac{A^{2}\beta'f\tilde{V}T}{2\psi}\tilde{V}
'-\frac{3M_{P}^{2}}{2}(\frac{ff'f''-f'^{3}}{f^{3}})\dot{T}^{2}$$
\begin{eqnarray}
-\left[\frac{(1-\beta')(\alpha'-2\beta')}{\beta'}\frac{f'}{f}-\frac{\alpha'}{T}
\right]\frac{\psi\dot{T}^{2}}{T},
\end{eqnarray}
\begin{eqnarray}
\ddot{T}+3H\dot{T}=\frac{2\left[(\frac{ff''+\beta'f'}{f^{2}})T\dot{T}^{2}-2
(\alpha'-2\beta'\frac{f'}{f}T)H\dot{T}\right]}{1+\frac{2\alpha'}{\beta'}-3\frac{f'}{f}T-
\frac{3M_{P}^{2}}{2}(\frac{f'}{f})^{2}\frac{T}{\psi}},
\end{eqnarray}

where $$\psi=\frac{\partial  \mathcal{L}}{\partial \Box
T}=-\frac{A\beta'\tilde{V}fT}{2h}$$ $$
h=\sqrt{1-(\alpha'f-2\beta'f'T)\partial_{\mu}T\partial^{\mu}T+\beta'fT
\Box T}$$ and $$\tilde{V}^{'}=\frac{d\tilde{V}}{dT}.$$
$H=\frac{\dot{a}}{a}$ is
the Hubble parameter.\\
The energy momentum tensor $T^{\mu\nu}$ is given by the standard
definition: $$\delta_{g_{\mu\nu}}S=-\int
d^{4}x\frac{\sqrt{-g}}{2}T^{\mu\nu}\delta g_{\mu\nu}.$$ So the
energy density, pressure and Friedman equation are found to be
\begin{eqnarray}
\rho=A\tilde{V}h+\frac{d}{a^{3}dt}(a^{3}\psi\dot{T})+(\alpha'f-2\beta'f'T)
\frac{A\tilde{V}}{h}\dot{T}^{2}-2\dot{\psi}\dot{T}+\frac{3M_{P}^{2}}{4}(\frac{f'}{f}^{2})\dot{T}^{2},
\end{eqnarray}
\begin{eqnarray}
p=-A\tilde{V}h-\frac{d}{a^{3}dt}(a^{3}\psi\dot{T})+\frac{3M_{P}^{2}}{4}(\frac{f'}{f}^{2})\dot{T}^{2},
\end{eqnarray}
\begin{eqnarray}
H^{2}=\frac{A}{3M_{P}^{2}}\tilde{V}h+\frac{d}{3M_{P}^{2}a^{3}dt}(a^{3}\psi\dot{T})+\frac{(\alpha'f-2\beta'f'T)}
{3M_{P}^{2}}\frac{A\tilde{V}\dot{T}^{2}}{h}
-\frac{2}{3M_{P}^{2}}\dot{\psi}\dot{T}+\frac{1}{4}(\frac{f'}{f}^{2})\dot{T}^{2}.
\end{eqnarray}
From Eq.(\ref{boun}), one can see that a successful bounce requires:
\begin{eqnarray}
\label{con} (\alpha'f-2\beta'f'T)
\frac{A\tilde{V}}{2hM_{P}^{2}}\dot{T}^{2}-\frac{\dot{\psi}\dot{T}}{M_{P}^{2}}+
\frac{3}{4}(\frac{f'}{f}^{2})\dot{T}^{2} <0.
\end{eqnarray}
When provided with a potential $\tilde{V}(T)$ from which to
construct an inflationary model, the slow roll approximation is
normally advertised as requiring the smallness of the two
parameters, denote by  $\epsilon$ and $\eta$, \cite{lid}.  With slow
roll approximation, equations (5) and (9) can be rewritten as,
\begin{eqnarray}
3H\dot{\psi}-\frac{A}{2}\tilde{V}'(T)\simeq0
\end{eqnarray}

\begin{eqnarray}
H^{2}\simeq\frac{A}{3M_{P}^{2}}\tilde{V}(T).
\end{eqnarray}
From equations (1), (7) and (8) one can obtain,
\begin{eqnarray}
\dot{H}=-\frac{1}{2M_{P}^{2}}\left[(\alpha'f-2\beta'f'T)
\frac{A\tilde{V}}{h}\dot{T}^{2}-2\dot{\psi}\dot{T}+
\frac{3M_{P}^{2}}{2}(\frac{f'}{f}^{2})\dot{T}^{2}\right],
\end{eqnarray}
so by using definition of $\psi$ and following assumption,
\begin{eqnarray}
\frac{3M_{P}^{2}}{2}\frac{f'^{2}}{f^{2}}\frac{1}{A\beta'f\tilde{V}}\ll1
\end{eqnarray}
The slow roll parameters are found to be
\begin{eqnarray}
\epsilon_{1}=-\frac{\dot{H}}{H^{2}}=\frac{M_{P}^{2}}{2A\beta'}\frac{\tilde{V}'^{2}\left(
f(1+\alpha)-f'T(2\beta'+1)+\frac{\beta'\tilde{V}'}{\tilde{V}fT}\right)}{\tilde{V}u^{2}}
\end{eqnarray}
\begin{eqnarray}
\epsilon_{2}=\frac{M_{P}^{2}}{A}\left[\frac{u'\tilde{V}'}{\tilde{V}u}+3\frac{\tilde{V}'^{2}}{u\tilde{V}^{2}}
-2\frac{\tilde{V}''}{\tilde{V}u}\right]
\end{eqnarray}
where $u=(\tilde{V}'fT+f'\tilde{V}T+f\tilde{V})$, then the  usual
slow roll parameters are $\epsilon=\epsilon_{1}$ and
$\eta=2\epsilon_{1}-\frac{1}{2}\epsilon_{2}$. Slow roll conditions
are $\epsilon_{1}\ll 1$ and $\mid \epsilon_{2}\mid\ll 1$. The true
end point of inflation, occurs at $\mid \epsilon_{2}(T_f)\mid\ll
\simeq 1$, where $T_f$ is the value of the tachyon at the end of
inflation. Inflation is commonly characterised by the number of
e-foldings of physical expansion that occur, this can be expressed
as \cite{lid}
\begin{eqnarray}\label{efol}
N=\ln
\frac{a_f}{a_i}=\frac{2A}{M_p^{2}}\int_{T_i}^{T_f}\frac{\tilde{V}}{\tilde{V'}}dT
\end{eqnarray}
Now we discuss on the stability of the model. The sound speed
express the phase velocity of the inhomogeneous perturbations of the
tachyon field \cite{c22}. To achieve the classical stability, we
must have $c_{s}^{2}\geq0,$ where
\begin{eqnarray}
c_{s}^{2}=\frac{p'}{\rho'}
\end{eqnarray}
where a prime denotes a derivative with respect to $T$. Therefore we
must have $\frac{p'}{\rho'}\geq0$. If $\dot{T}=0$, then eqs. (7),
(8) leads to the following equations respectively
\begin{eqnarray}
\rho'=Ah(\frac{V'}{f^{2}}-\frac{2f'V}{f^{3}})+\frac{AV}{f^{2}}h'+\frac{1}{a^{3}}\frac{d}{dT}(a^{3}\psi
\ddot{T})
\end{eqnarray}
\begin{eqnarray}
p'=Ah(\frac{2f'V}{f^{3}}-\frac{V'}{f^{2}})-\frac{AV}{f^{2}}h'+\frac{1}{a^{3}}\frac{d}{dT}(a^{3}\psi
\ddot{T}).
\end{eqnarray}
If $p'=0$, then (20) gives,
\begin{eqnarray}
Ah(\frac{2f'V}{f^{3}}-\frac{V'}{f^{2}})=\frac{AV}{f^{2}}h'+\frac{1}{a^{3}}\frac{d}{dT}(a^{3}\psi
\ddot{T}).
\end{eqnarray}
But if $p'\neq0$ , to have $\frac{p'}{\rho'}>0$, using eqs. (19),
(20), following condition must be satisfy
\begin{eqnarray}
\frac{1}{a^{3}}\frac{d}{dT}(a^{3}\psi \ddot{T})}>
-Ah(\frac{2f'V}{f^{3}}-\frac{V'}{f^{2}})+{\frac{AV}{f^{2}}h'
\end{eqnarray}
On the other hand, if $\dot{T}\neq0$, to achieve to the classical
stability following conditions must be satisfy respectively for the
cases $p'=0$ and $p'\neq0$:
\begin{eqnarray}
Ah(\frac{2f'V}{f^{3}}-\frac{V'}{f^{2}})=\frac{AV}{f^{2}}h'+\frac{1}{a^{3}}\frac{d}{dt}(a^{3}\psi'
\dot{T})+
\frac{3M_{P}^{2}}{4}\frac{f'}{f}\dot{T}^{2}(\frac{f'^{2}}{f}-2f''),
\end{eqnarray}
\begin{eqnarray}
\frac{p}{p+2A\tilde{V}h+2\frac{d}{a^{3}dt}(a^{3}\psi\dot{T})+(\alpha'f-2\beta'f'T)
\frac{A\tilde{V}}{h}\dot{T}^{2}-2\dot{\psi}\dot{T}}>0.
\end{eqnarray}

 We will show below that
Eq.(\ref{con}) can be satisfied easily
for our model.\\
In Fig.1, we take $V(T)=V_{0}e^{-\lambda T^{2}}$(motivated by string
theory) and $f(T)=1+\sum_{i=1}c_{i}T^{2i}$. One can see from this
figure the EoS crosses $\omega=-1$, which gives rise to a possible
inflationary phase after the bouncing.\\
In Fig.2 we take a different potential
$V(T)=\frac{V_{0}}{e^{\lambda\phi+e^{-\lambda\phi}}}$ for numerical
calculations. This figure show the crossing over -1 for EoS again.
Also the Fig.1 and Fig.2 show that the Hubble parameter $H$ running
across zero at t=0 which we have choosed it, as the bouncing
point.\\

\begin{tabular*}{2cm}{cc}
\hspace{0.25cm}\includegraphics[scale=0.25]{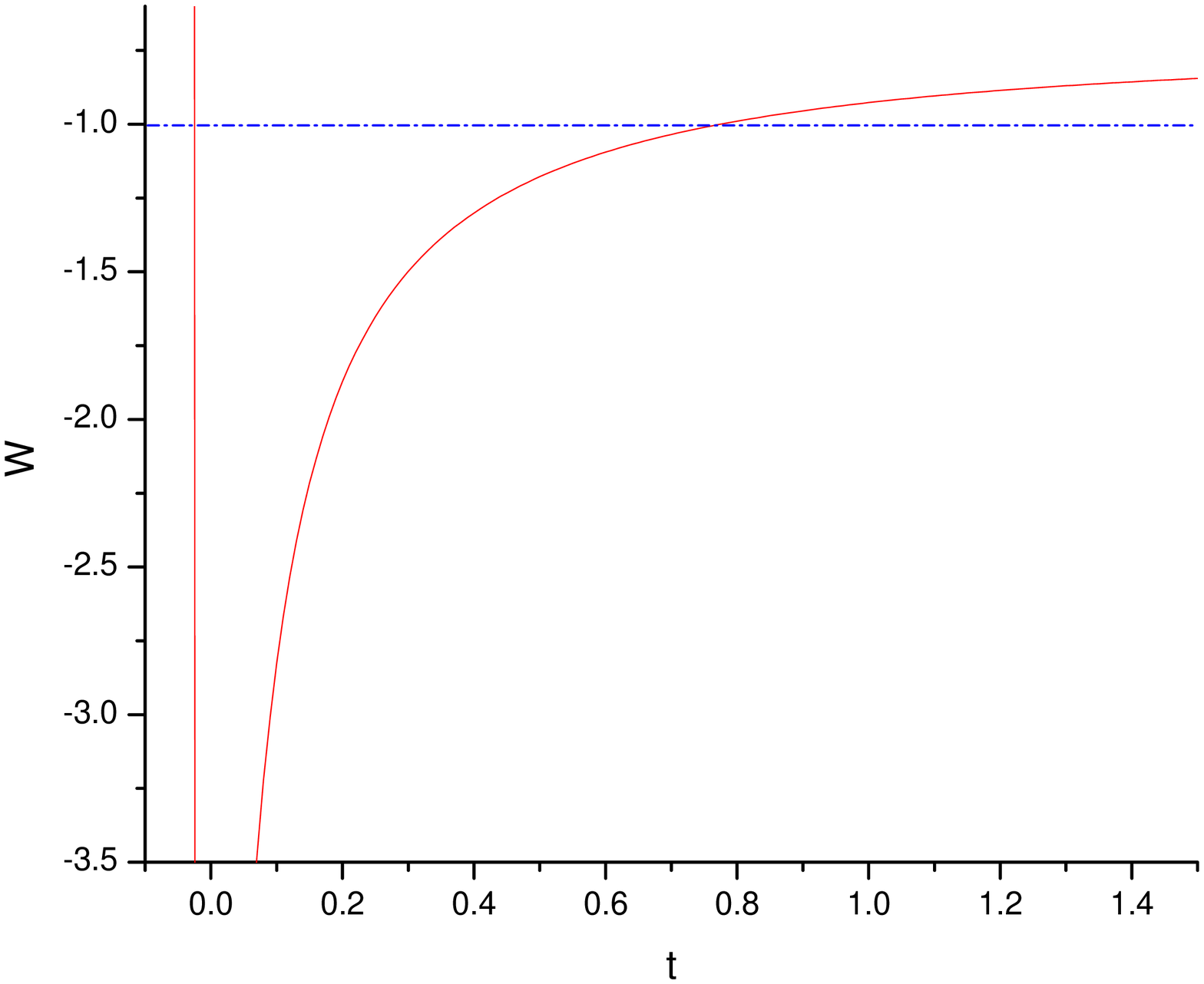}\hspace{0.5cm}\includegraphics[scale=0.25]{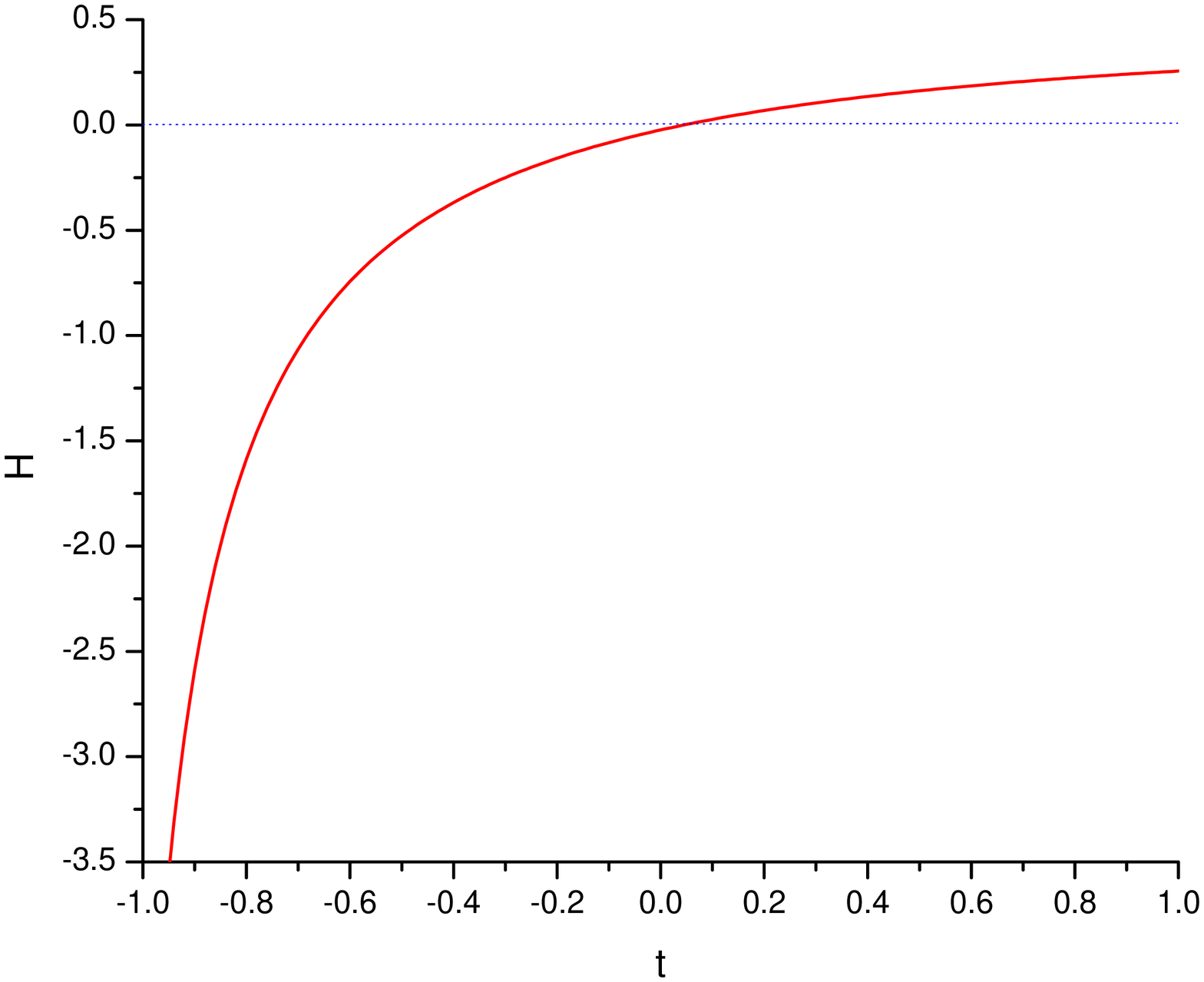}\hspace{0.5cm}\\
\hspace{2 cm}\textbf{Figure 1:} \,The plot of EoS and the Hubble
parameter for the potential $V(T)=e^{-\lambda T^{2}}$,\\~~~~~~~~~~
 $\alpha=-2$ and $\beta=2.2$. Initial values are $\phi=1$ and $\dot{\phi}=3$. \\
\end{tabular*}\\\\\\
\begin{tabular*}{2cm}{cc}
\hspace{0.25cm}\includegraphics[scale=0.25]{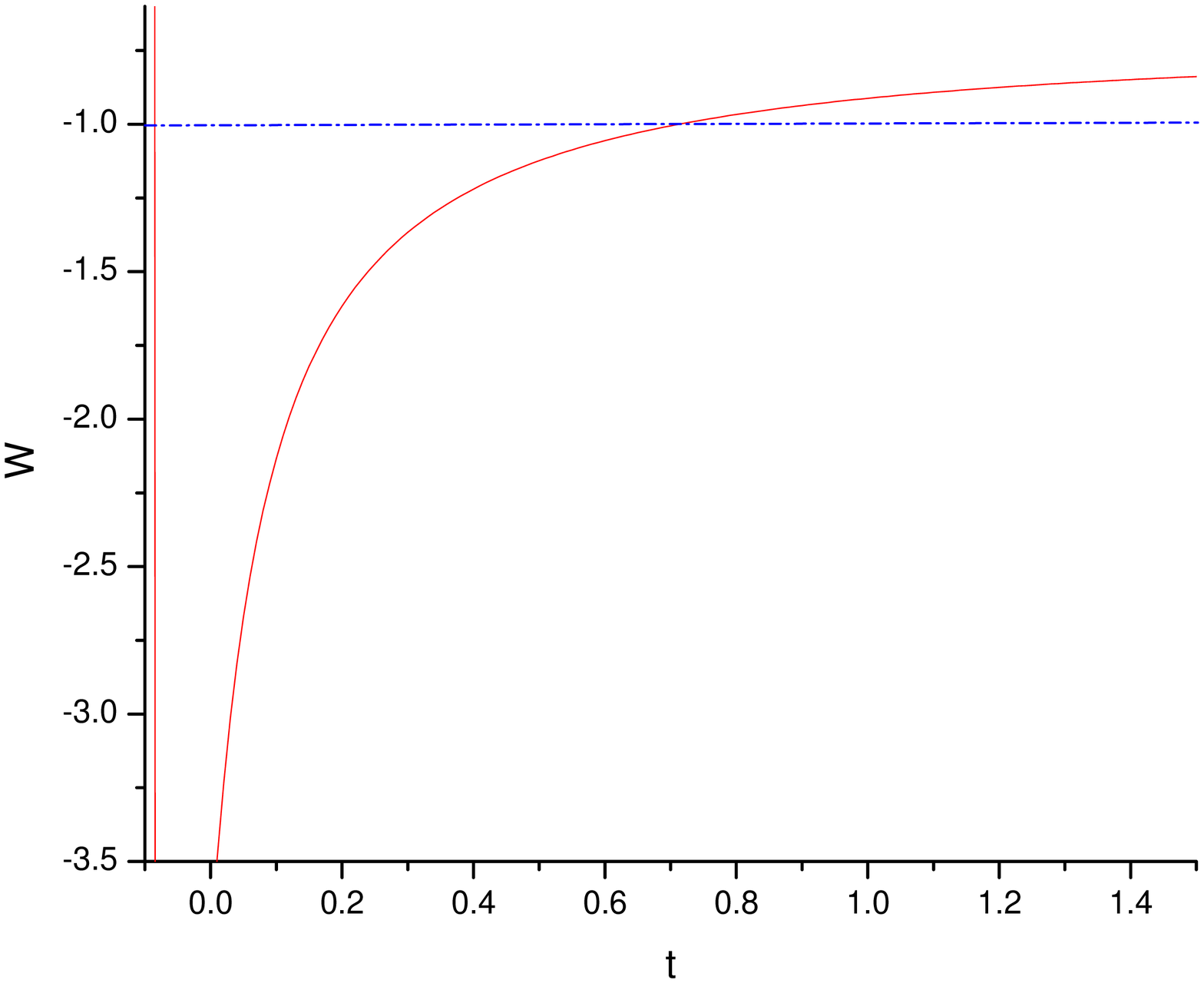}\hspace{0.5cm}\includegraphics[scale=0.25]{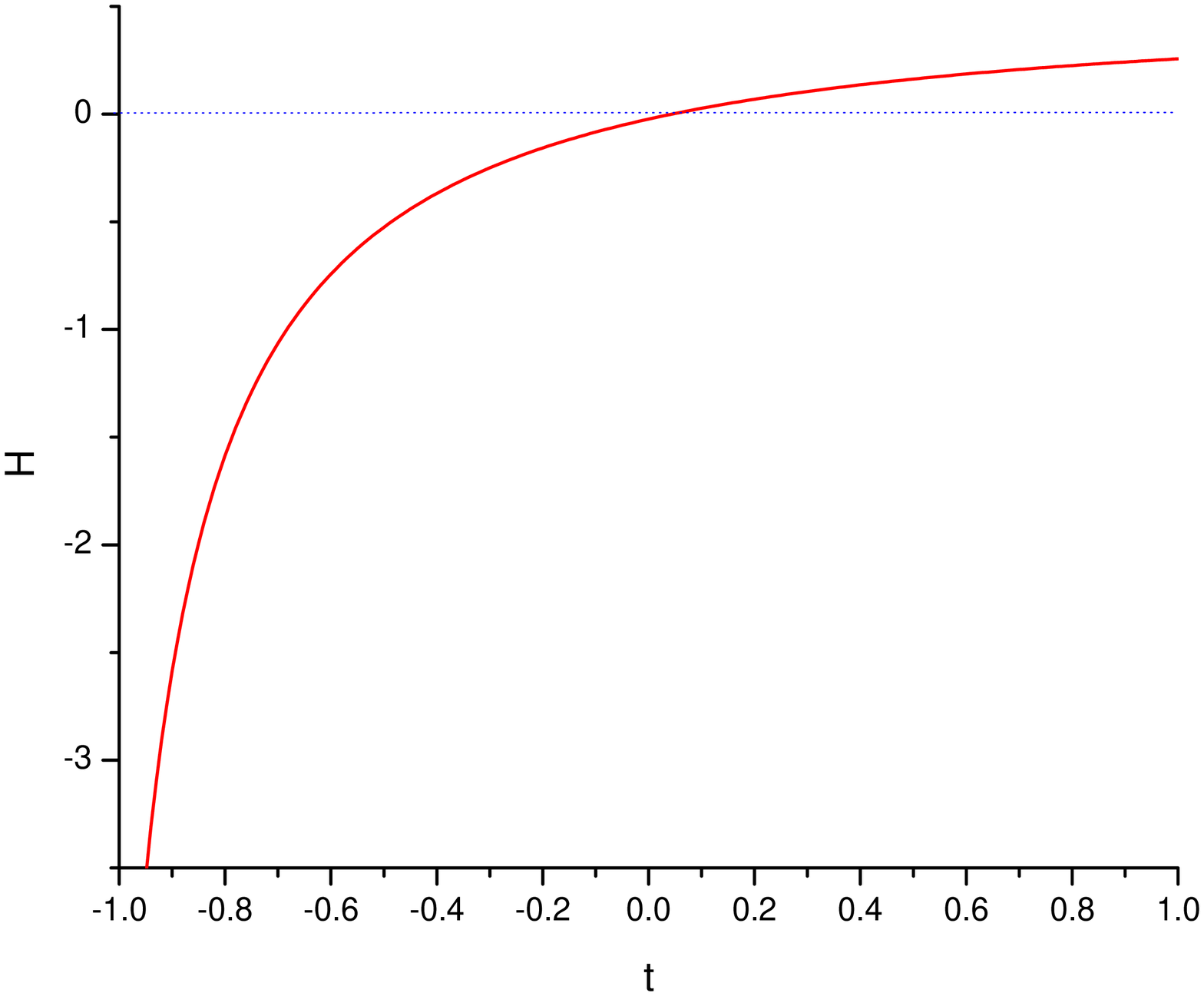}\hspace{0.5cm}\\
\hspace{1.7cm}\textbf{Figure 2:} \,The plot of EoS and the Hubble
parameter for the potential $V(T)=\frac{V_{0}}{e^{\lambda
T}+e^{-\lambda
T}}$,\\~~~~~~~~~~~~~$\alpha=-2$, $\beta=2.2$, $\lambda=2$ and $V_{0}=5$. Initial values are $\phi=1$ and $\dot{\phi}=3$. \\
\\
\end{tabular*}\\\\\\
In order to discuss if our bouncing model have the event and/ or
particle horizons we need to obtain the Hubble parameter. The
particle horizon, $R_p$, is given by
\begin{eqnarray}
R_p=a\int_0^t\frac{dt}{a}=a\int_0^a\frac{da}{Ha^2}.
\end{eqnarray}
On the other hand the event horizon, $R_h$, is as
\begin{eqnarray}
R_h= a\int_t^\infty \frac{dt}{a}=a\int_a^\infty\frac{da}{Ha^2}
\end{eqnarray}
We see that equations (7), (8) and (9) are very complicated to give
us  the explicit form of  $H(a)$. But by numerical calculation in
equations (7), (8) we have shown   $\omega$ across -1, so one can
obtain the
 form of $\omega$ approximately.  In our model $H(a)$  can be obtained  by  solving Friedman equation
 and  some numerical calculations. The following $H(a)$  completely
 support the $\omega$ across -1,
\begin{equation}
H(a)=\frac{\sqrt{(1-q)a^{3(1-q)}-1}}{(1-q)^{\frac{3}{2}}a^{3(1-q)}},
\end{equation}
where $q<1.$
 For instance in our two examples by both tachyon potentials we
 have $q\simeq0.51$.\\
 If we take the above result for $H(a)$ in the above formula for
 event horizon, the answer will diverge as $a\rightarrow \infty$.
 So we can't see  the event horizon in our model but we have particle
 horizon.
\section{Conclusion}
In this letter, we have studied the bouncing solution in the
universe dominated by the quintom matter. We have assumed that the
quintom matter non-minimally coupled to gravity with an extra term
in the usual effective action of tachyon dynamics. By performing a
conformal transformation we have obtained the new action. In order
to discuss the bouncing solution we have derived the corresponding
energy density pressure and  Friedman equation for this model. Also
we have obtained the bouncing condition as Eq.(\ref{con}), then by
considering a couple example for potential of scalar filed we have
shown that the mentioned condition can be satisfy. Then we have
obtained the slow roll parameters in terms of tachyon potential,
$f(T)$, and another parameters of models. Also we have obtained the
e-folding as Eq.(17). After that we have studied the conditions for
the classical stability in our interesting model. Generally we know
that the condition for classical stability is given by
$c_{s}^{2}\geq0, $\cite{c22}, where  $c_{s}$ is the sound speed
express the phase velocity of the inhomogeneous perturbations of the
tachyon field. To achieve the classical stability, our model must
satisfy conditions, (21) or (22), and (23) or(24) respectively for
$\dot{T}=0$ and $\dot{T}\neq0$. Finally we have shown that such
bouncing model has not event horizon, but has particle horizon.
 
\end{document}